\documentclass[11pt]{article}
\usepackage{moriondav,epsfig}

\def\Journal#1#2#3#4{{#1} {\bf #2}, #3 (#4)}

\def\PLB{{\em Phys. Lett.}  B}
\def\PRL{\em Phys. Rev. Lett.}
\def\PRD{{\em Phys. Rev.} D}

\def\ra{\rightarrow}

\newcommand{\met}     {\mbox{$\not\!\!{E_T}$}}
\def\gsim{\mathrel{\rlap{\raise.4ex\hbox{$>$}} {\lower.6ex\hbox{$\sim$}}}}
\def\lsim{\mathrel{\rlap{\raise.4ex\hbox{$<$}} {\lower.6ex\hbox{$\sim$}}}}

\begin{document}
\vspace*{4cm}
\title{SEARCHES FOR THE STANDARD MODEL HIGGS AT THE TEVATRON}

\author{ AVTO KHARCHILAVA \\
(For the CDF and D\O\ Collaborations) }

\address{Department of Physics, 225 Nieuwland Science Hall,
Notre Dame, IN 46556, USA}

\maketitle\abstracts{
Preliminary results obtained by the CDF and D\O\ Collaborations on the
Standard Model Higgs searches at the Tevatron Run~II are discussed.
The data that correspond to an integrated luminosity of
$\gsim$~150 pb$^{-1}$ are compared to theoretical expectations.
Various final states involving the $W$ and $Z$ bosons production
are examined, and new limits are set on the
$\sigma[W(\rightarrow e\nu/\mu\nu)H(\rightarrow b\bar{b})]$ and
$\sigma[H(\rightarrow WW^{(*)})]$ cross sections.}

\section{Introduction}

One of the main physics objectives at the upgraded Tevatron Run II
is to search for the Standard Model (SM) Higgs boson. Prospects for
the discovery are not slim thanks to the recent improvements in
the accelerator performance, and due to indirect
constraints on the mass that favor a light Higgs boson
of 117~GeV with an upper limit at 251~GeV~\cite{fit_h_mass}.
The direct searches at LEP have already excluded the SM Higgs boson below
114.4~GeV~\cite{lep_h_mass}.

In this range of masses,
the Higgs boson production cross sections at Tevatron energies are small,
of the order of 1~-~0.1~pb depending on its mass and the
production mechanism. The rate is largely dominated by the gluon fusion,
$gg \ra H$.
At low masses ($m_H \lsim 135$ GeV) however, when $H \ra b\bar{b}$ decays
are dominant,
this mechanism suffers from the overwhelming QCD background. Instead,
the $WH$ and $ZH$ associated production can be explored with vector bosons
decaying leptonically to handle the background.
At higher masses, $m_H \gsim 135$ GeV,
the $H\rightarrow WW^{(*)}\rightarrow\ell\nu\ell\nu \; (\ell = e, \mu)$ decays
are promising to look at, although the Higgs boson cannot be
reconstructed explicitly due to presence of neutrinos in the final states.

While to discover, or rule out, the SM Higgs boson at the Tevatron would
require several fb$^{-1}$ of data~\cite{run2_prospects}, the 
modest luminosities of $\gsim 150$~pb$^{-1}$ that have been
collected and analyzed so far by CDF and D\O\
allow to optimize detector performance, study the background processes,
and extend the previous limits on the Higgs boson production.

\section{$WH$ associated production}

\begin{figure}
\centerline{
\epsfig{figure=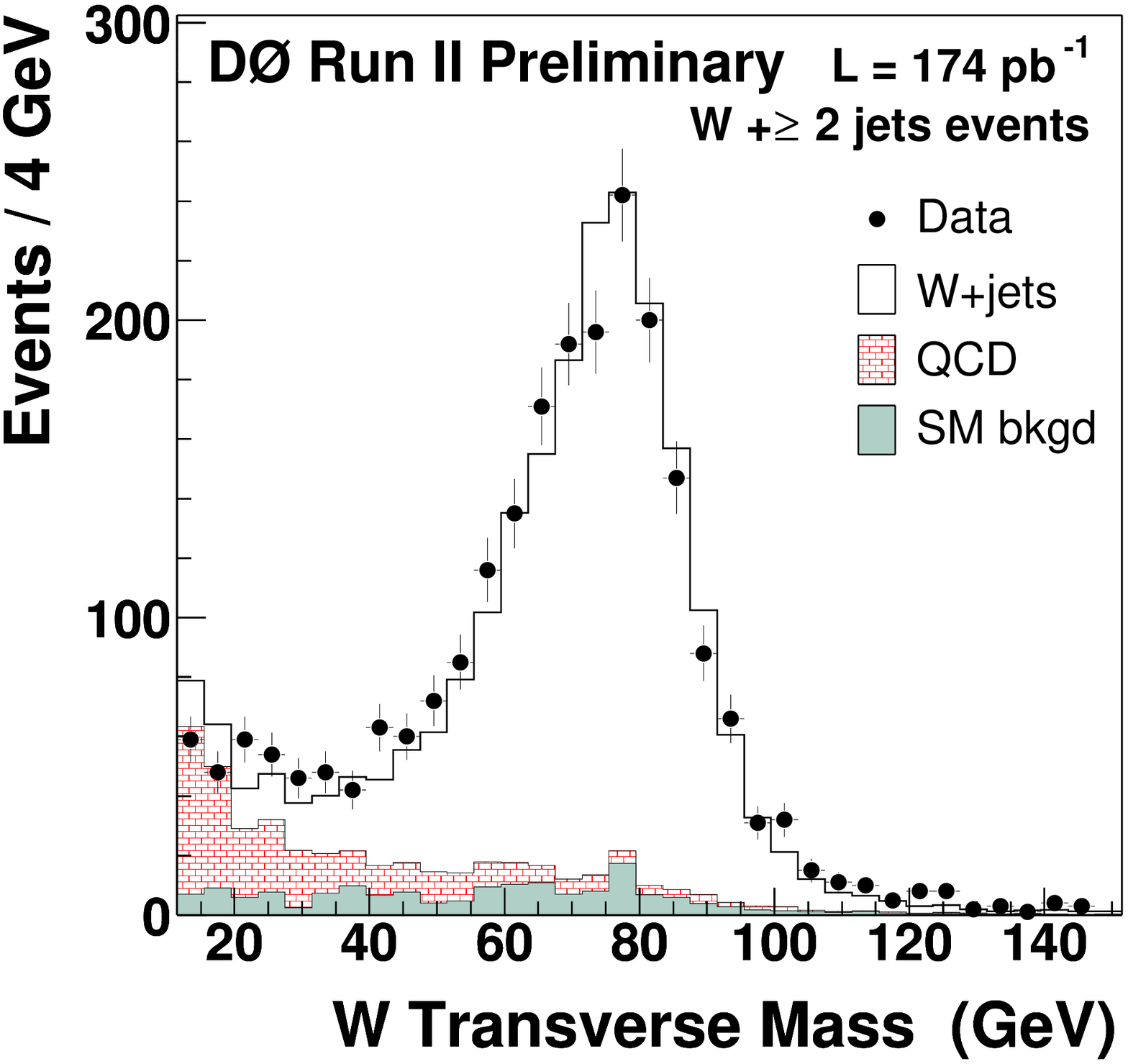,height=50mm}
\hskip -0.2cm
\epsfig{figure=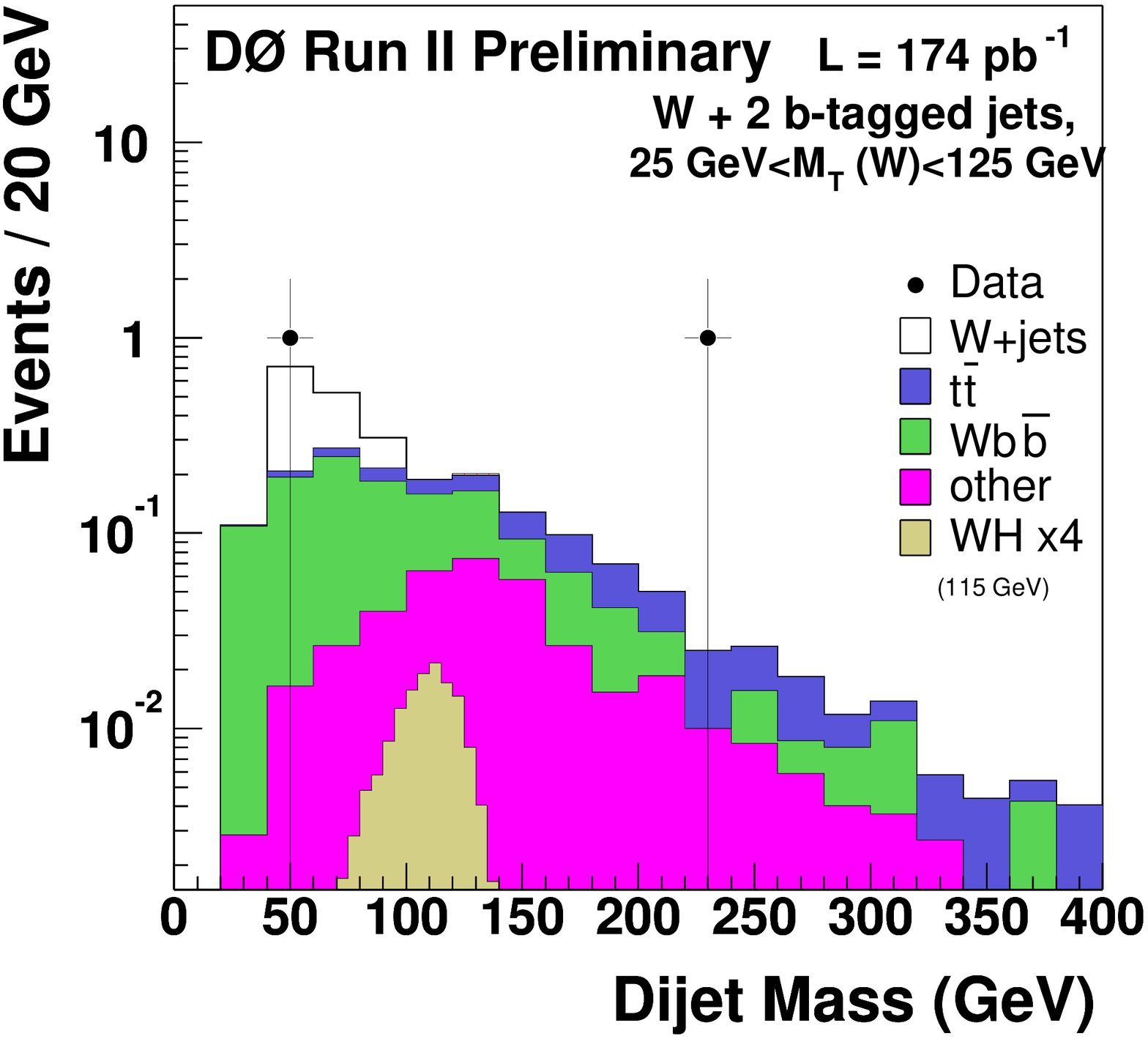,height=50mm}
\hskip -0.3cm
\epsfig{figure=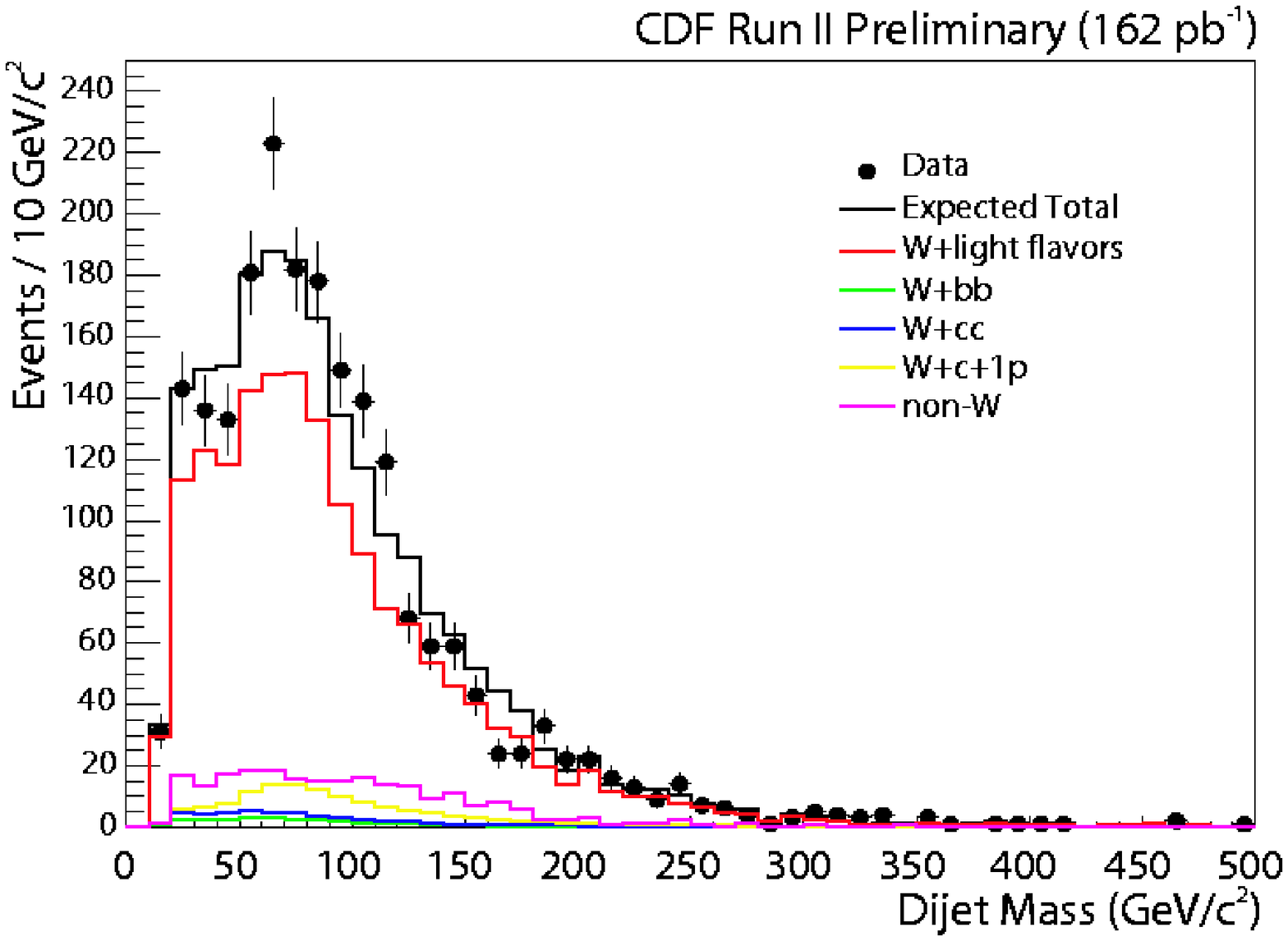,height=50mm}
}
\vskip -45mm
{\hskip 3mm} a) {\hskip 47mm} b) {\hskip 98mm} c)
\vskip 38mm
\caption{a) The $W$ transverse mass distribution obtained
by D\O\ in $e+\met+jets$ final states. b) The di-jet mass spectrum
when both jets are tagged as b's by D\O; light gray (brown) histogram
corresponds to 4 times the expected Higgs signal of 120 GeV.
c) Un-tagged di-jet mass distribution in $e/\mu+\met+jets$ final states
by CDF.
\label{fig:cdfd0_smhiggs1}}
\end{figure}

The first step of the
search strategy is to identify the $W$ production through
its decay to a charged lepton and a neutrino. In addition, two jets
in the event are required with at least one of them tagged as a b-jet.
D\O\ has analyzed 174 pb$^{-1}$ of data and observed nearly 2,600
$W(\rightarrow e\nu) + \geq 2jets$ candidates. These events are
selected by requiring a central, isolated electron with
$p_T > 20$~GeV, missing transverse energy of $\met > 25$~GeV and jets with
$p_T > 20$~GeV in a pseudorapidity region $|\eta| < 2.5$.
A good agreement between data and MC has been obtained for various kinematic
properties of the event. An example of the $W$ transverse mass distribution
is shown in Fig.~\ref{fig:cdfd0_smhiggs1}a. The $W+2jets$
processes are simulated by ALPGEN~\cite{alpgen} parton generator interfaced
with {\sc Pythia}~\cite{pythia}, but the cross sections are normalized to
NLO calculations~\cite{nlo_w2jets}.
The QCD background is evaluated from the data,
while other SM processes are generated with {\sc Pythia}.
Only two jet events are retained in the analysis
(to suppress the $t\bar{t}$ background) with both tagged as b-jets.
Figure~\ref{fig:cdfd0_smhiggs1}b shows the di-jet invariant mass distribution
in the final sample. Two events are observed in data while $2.5\pm0.5$ events
are expected from simulations. The background is dominated by the $Wb\bar{b}$
production ($1.4\pm0.4$) with a non-negligible contribution from
the $Wc\bar{c}/jj$ ($0.4\pm0.3$), $t\bar{t}$ and single top
processes ($0.6\pm0.2$). An upper limit on the $WH$ associated production
cross section,
$\sigma[W(\rightarrow e \nu)H(\rightarrow b\bar{b})]~<~12.4$~pb,
has been set at $95\%$ C.L. by D\O\  for
a Higgs mass of 115~GeV.

\begin{figure}
\centerline{
\epsfig{figure=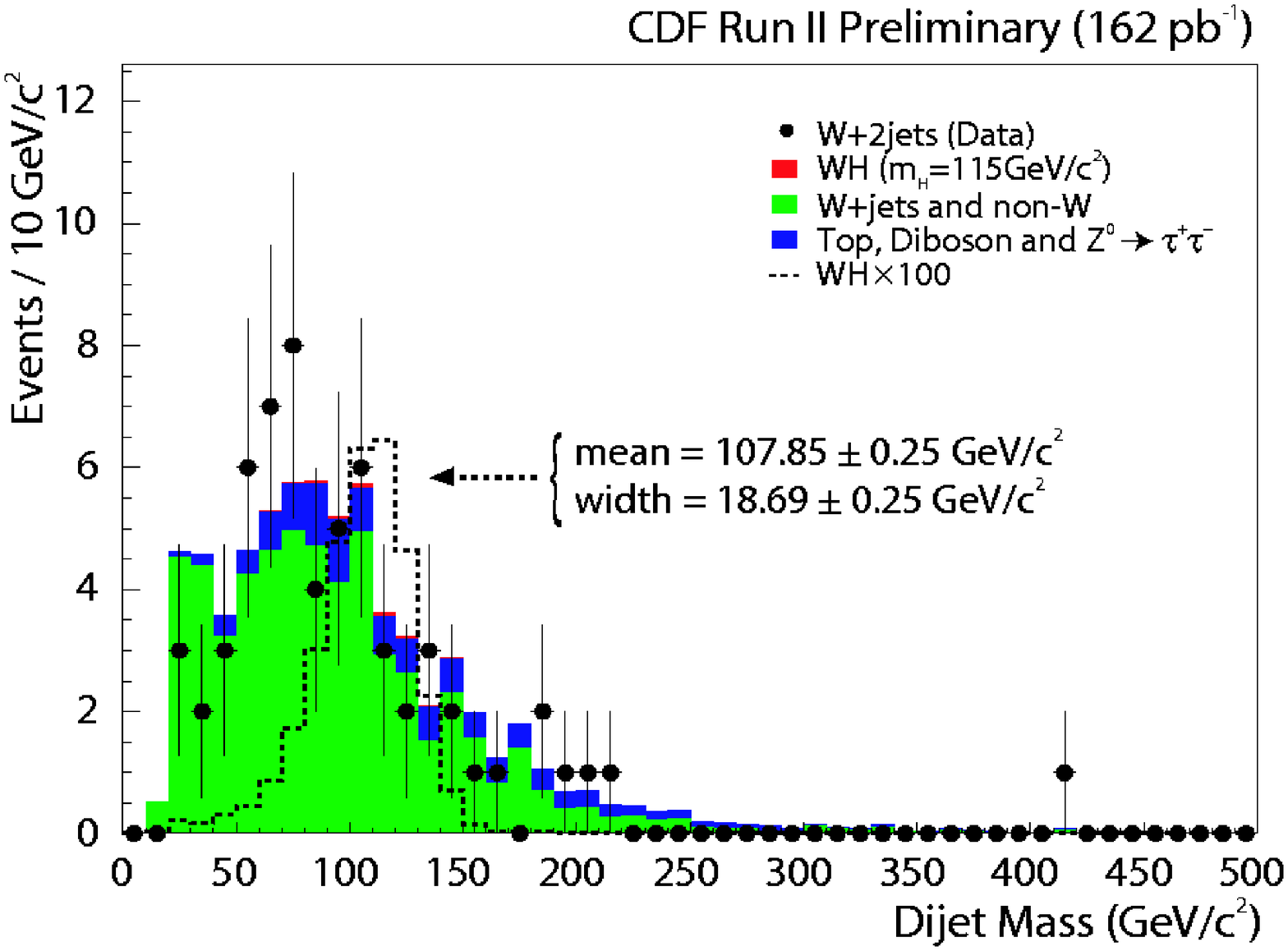,height=55mm}
\hskip 0.0cm
\epsfig{figure=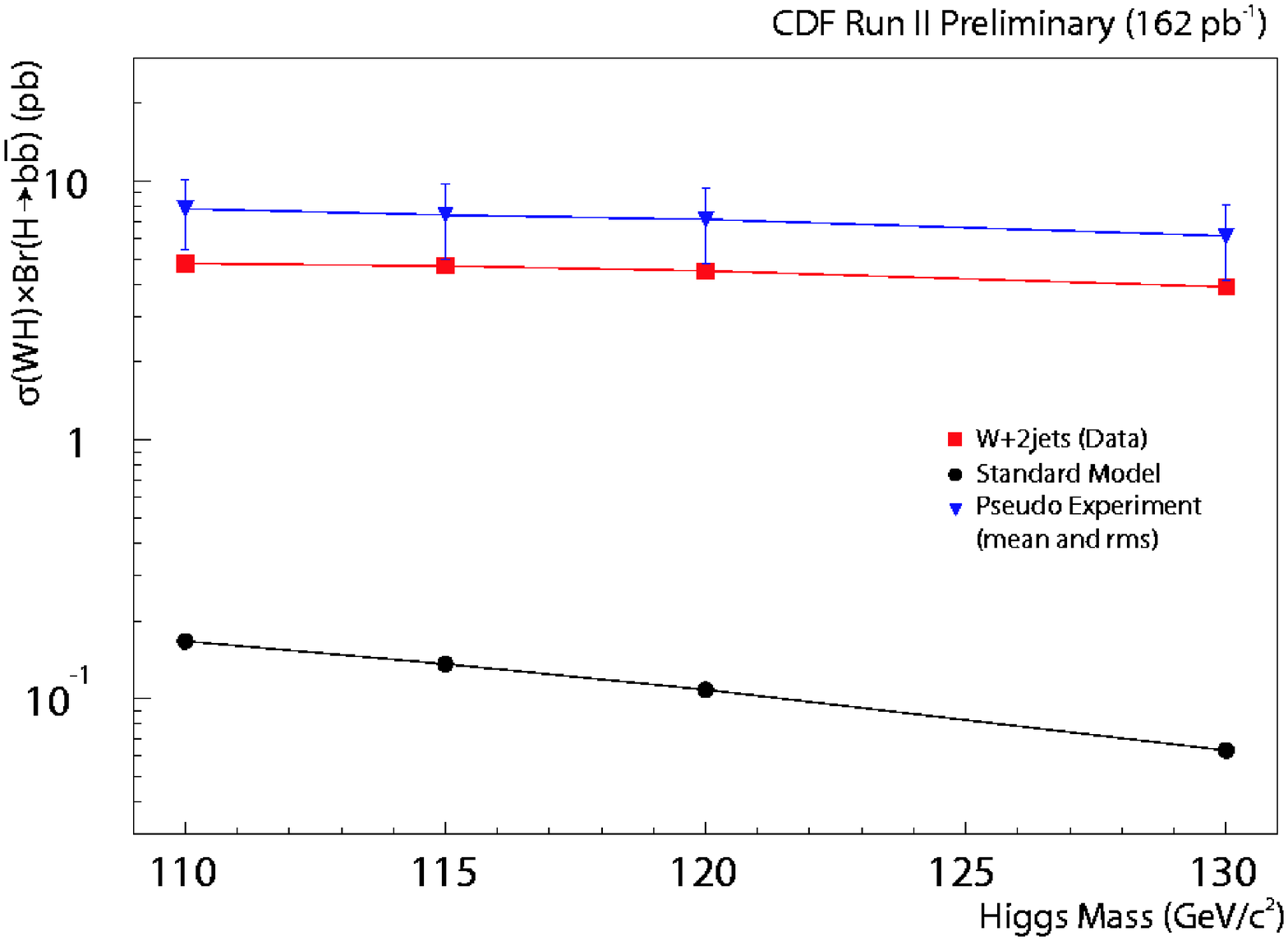,height=55mm}
}
\vskip -51mm
{\hskip 14mm} a) {\hskip 70mm} b)
\vskip 44mm
\caption{a) Di-jet mass spectrum in $\ell+\met+jets$ events with
at least one b-jet obtained by CDF.
b) Limit on the $WH$ production cross section as function of the
Higgs boson mass; dots indicate the SM expectations.
\label{fig:cdfd0_smhiggs2}}
\end{figure}

CDF uses both, electron and muon decay modes of $W$, to effectively
double the size of the initial data sample which corresponds to an integrated
luminosity of 162~pb$^{-1}$. The di-jet sample
is selected by requiring central, isolated electron/muon with
$p_T > 20$~GeV, $\met > 20$~GeV, two jets with $E_T > 15$~GeV in $|\eta|<2$,
and vetoing extra jets or isolated leptons in an event.
Figure~\ref{fig:cdfd0_smhiggs1}c compares the di-jet mass distribution
in data (about 2,100 events in total) with ALPGEN plus HERWIG~\cite{herwig}
simulations. At least one b-jet is then
required that leaves 62 events in data with $61\pm5$ events expected in MC.
The background is dominated by fake tags in $Wjj$ events (14),
contribution from $Wc\bar{c}/b\bar{b}$ (13/12),
QCD (10) and top production (9).
Figure~\ref{fig:cdfd0_smhiggs2}a shows the di-jet mass
distribution in these events indicating a good agreement between
data and MC. It also shows simulated Higgs signal
with a mass of 115~GeV, but 100 times the expected rate. In the absence of
a signal, CDF sets upper limits on the $WH$ production,
$\sigma[W(\rightarrow e\nu/\mu \nu)H(\rightarrow b\bar{b})]~\lsim~5$~pb
at $95\%$~C.L.
that are shown in Fig.~\ref{fig:cdfd0_smhiggs2}b as function of the
Higgs boson mass. Both, CDF and D\O\ results discussed here
are superior to similar Run~I measurements~\cite{cdf_wh_runi}.

\section{Measurement of the $\sigma[Z+b$-$jet]/\sigma[Z+jet]$
cross section ratio}

The $Z$ boson production in association with b-jet(s) is an important
process to study for several reasons: it probes 
the b-quark content of the proton parton distribution functions, 
it is a dominant background to the $ZH$
associated production and a benchmark process for non-SM Higgs searches
in $bh/b\bar{b}h$ associated production.
D\O\ has analyzed electron and muon decays of $Z$ and presented
results on $\sigma[Z+b$-$jet]/\sigma[Z+jet]$, since in the ratio
many systematic uncertainties cancel.
The event selection requires two isolated leptons with
$p_T > 15$~GeV in $|\eta| < 2.5$ (electron channel) or
$p_T > 20$~GeV, $|\eta| < 2$ (muon channel). $Z$ peak is selected for
signal events while the side bands are used for the background evaluations.
Events are required to have a jet with
$E_T > 20$~GeV in $|\eta| < 2.5$.
Since the b-tagging algorithm cannot distinguish between b- and c-jets,
their relative content in $Z+jet$ sample is taken from the NLO
calculations~\cite{zbzj_theory}, while the total rate
(including light quarks/gluon jets) is normalized to data.
This leaves one unknown quantity to measure -- the fraction of events
with one b-tagged jet provided the (mis-)tagging
rates for light quarks and b-/c-jets are known; these rates
are in fact measured in data.

Preliminary result on the cross section ratio is
$0.024\pm0.005 (stat.)^{+0.005}_{-0.004} (syst.)$ as compared to $\sim$~0.02
predicted in theory~\cite{zbzj_theory}. The systematic errors include
uncertainties due to the jet tagging efficiency ($16\%$), the jet energy scale
($11\%$), background estimations ($6\%$) and assumtions on the
relative fraction of b- and c-jets ($3\%$). This measurement by D\O\
is the first of its kind.

\section{Higgs searches in
$H \rightarrow WW^{(*)}\rightarrow\ell^{+}\nu\ell^{-}\bar{\nu}$ final states}

At higher Higgs boson masses, $m_H \gsim 135$~GeV, when the
$WW^{(*)}$ decay mode becomes kinematically accessible, one can explore
leptonic decays of $W$ to handle various backgrounds.
The signal signature is: two isolated, opposite charge leptons
accompanied by a large missing $E_T$.
The background processes include
the $Z/\gamma^*(\rightarrow \ell^+\ell^-)+jets \; (\ell = e, \mu, \tau)$, $WW$,
$WZ$, $W+jets$, $t\bar{t}\rightarrow b\ell^+\nu \bar{b}\ell^-\bar{\nu}$
and QCD production.

\begin{figure}
\centerline{
\epsfig{figure=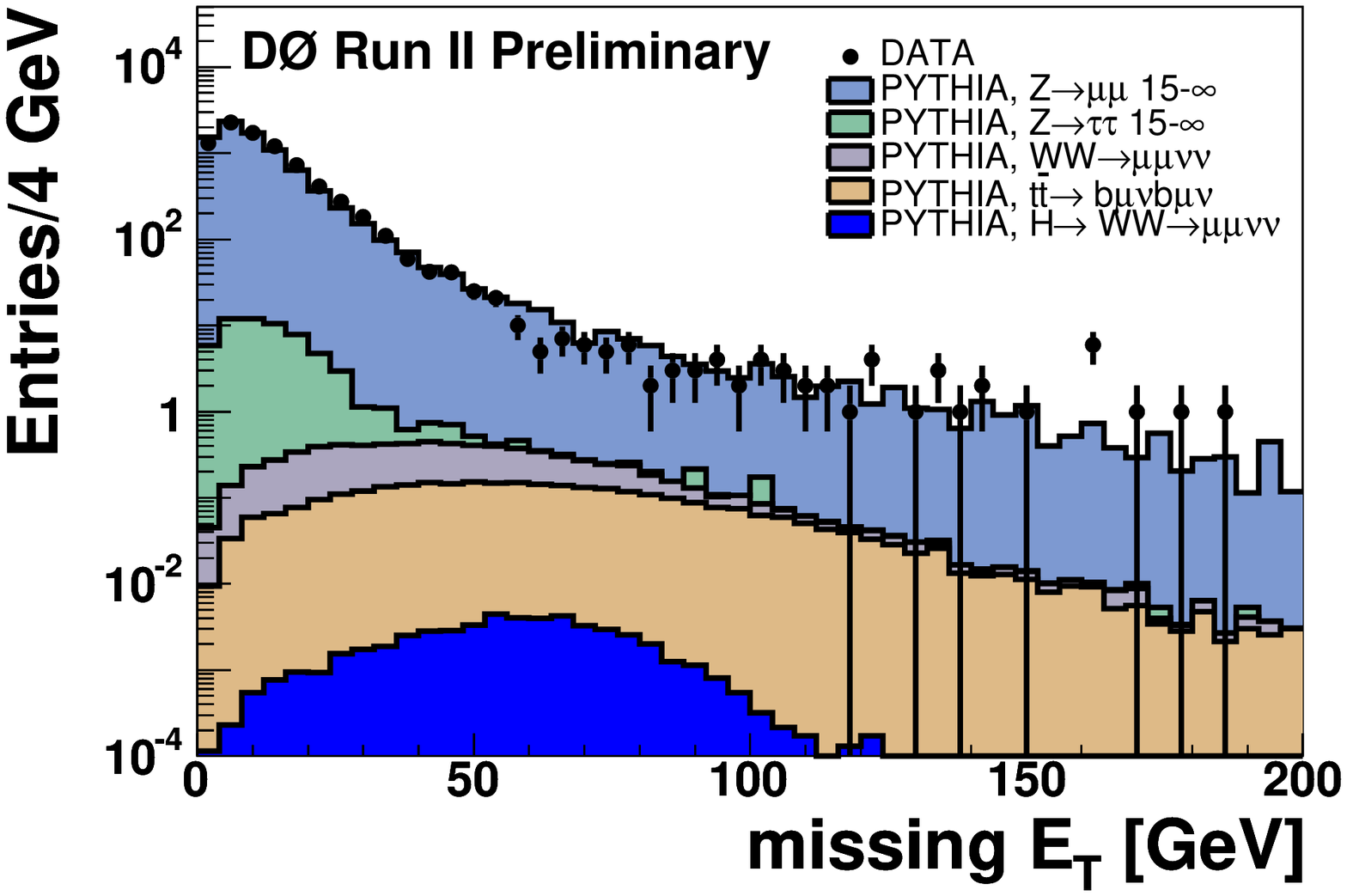,height=50mm,width=65mm}
\hskip -0.3cm
\epsfig{figure=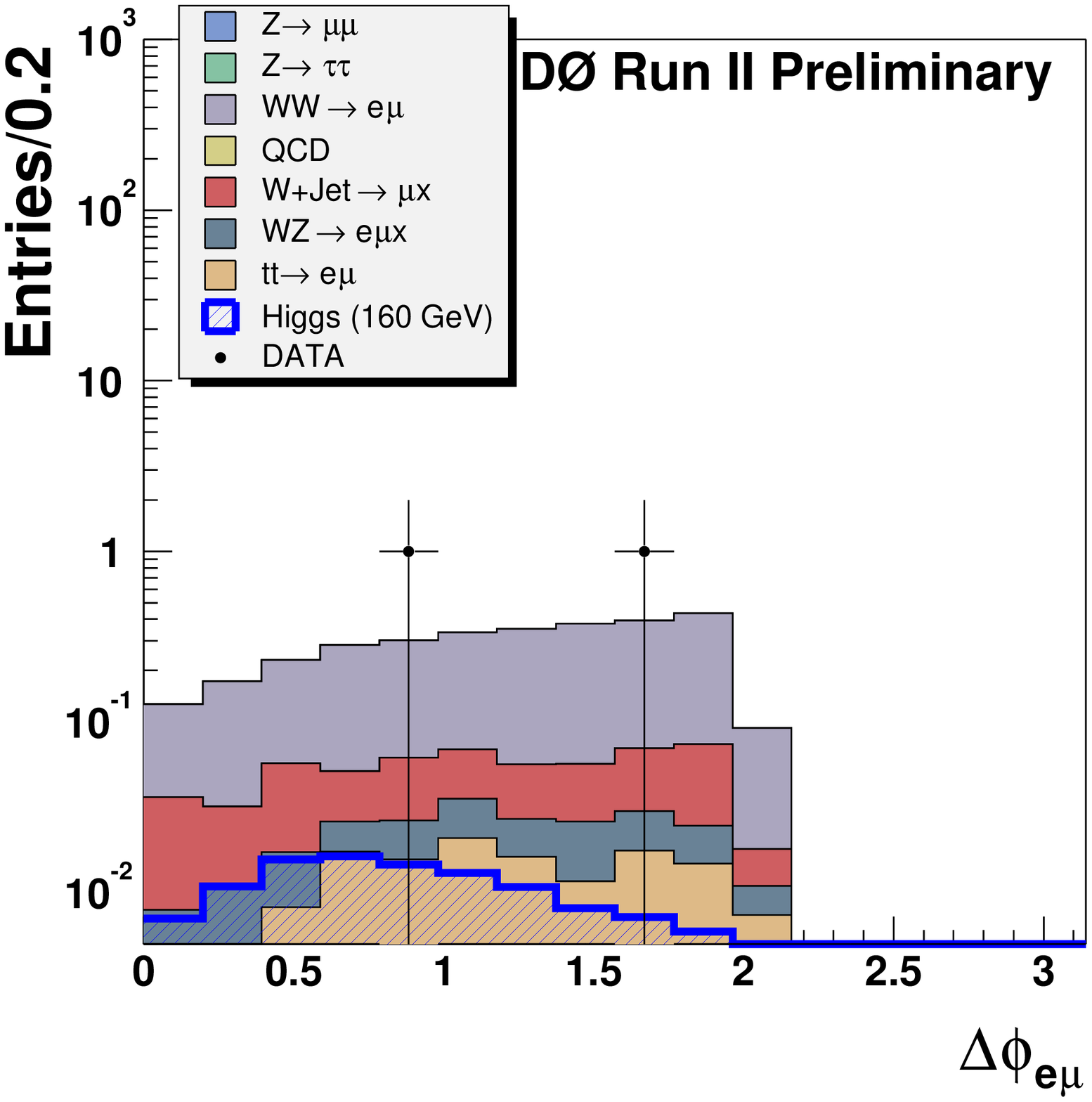,height=50mm}
\hskip -0.3cm
\epsfig{figure=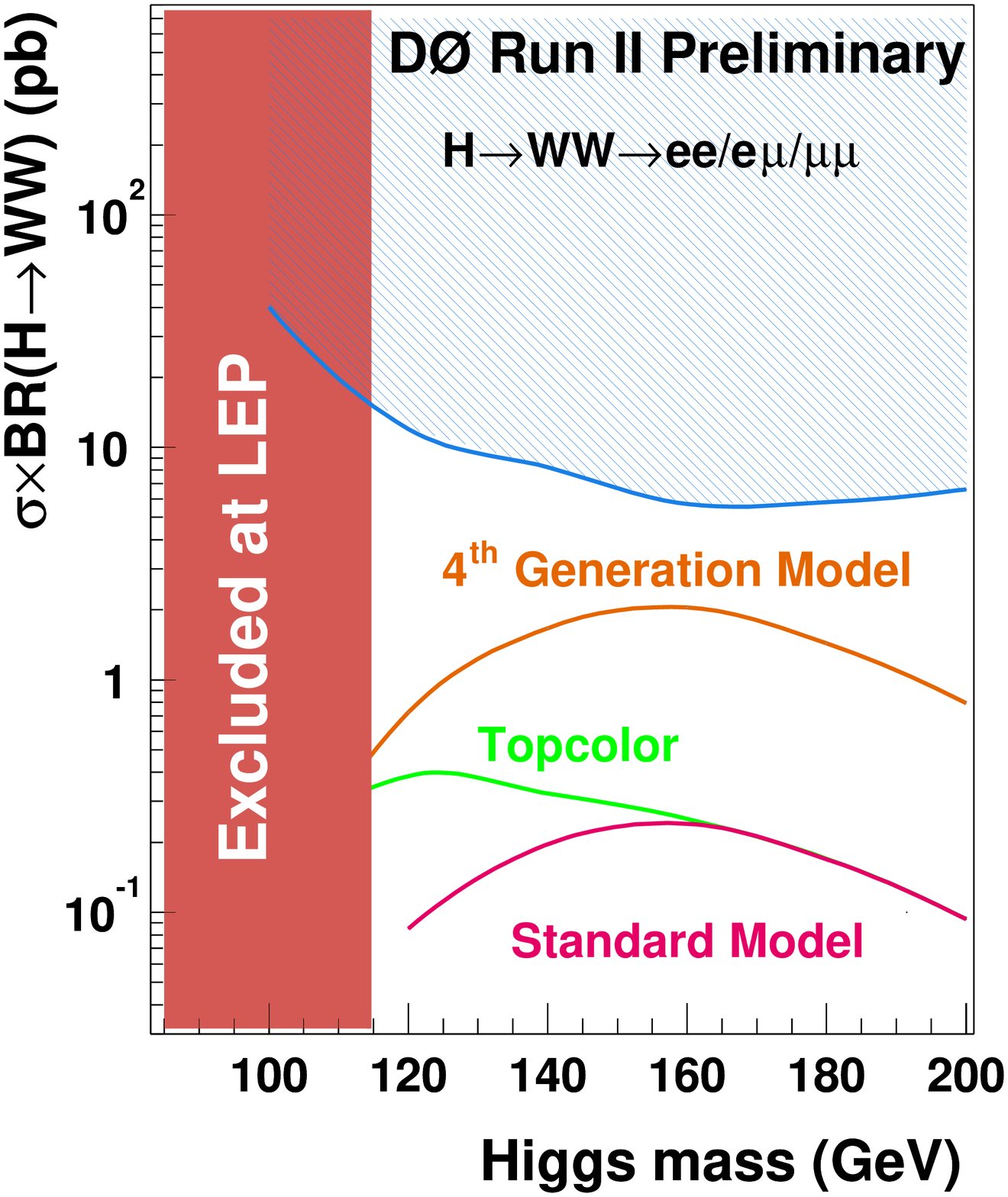,height=50mm,width=50mm}
}
\vskip -45mm
{\hskip 12mm} a) {\hskip 86mm} b) {\hskip 44mm} c)
\vskip 37mm
\caption{a) The missing transverse energy distribution in di-muon
sample after event pre-selection. b) Distribution of the azimuthal
opening angle between electron and muon; dark (blue) histograms in
both plots correspond to the Higgs signal of 160~GeV. c) Excluded cross
section times branching ratio to $W$-pairs,
$\sigma \times BR(H \rightarrow WW^{(*)})$, along with the expectations
from the SM Higgs boson production and alternative models.
\label{fig:d0_hww}}
\end{figure}

D\O\ has analyzed $\sim$~180, 160 and 150 pb$^{-1}$ of data in $ee$, $e\mu$ and
$\mu\mu$ final states, respectively.
Selected events must have two leptons with
$p_T > 12, \; 8$~GeV
($ee/e\mu$ channels), $p_T > 20, \; 10$~GeV ($\mu\mu$),
$\met > 20$~GeV ($ee/e\mu$),
$\met > 30$~GeV ($\mu\mu$), no $Z$ candidates and energetic jets.
Simulations are done with {\sc Pythia} except for the QCD contributions
which are evaluated from the data. A good agreement between data and MC
is obtained at each step of the event selection.
As an example, Fig.~\ref{fig:d0_hww}a shows
the $\met$ spectrum in di-muon final states after
event pre-selection ($p_T$ and isolation criteria).
Figure~\ref{fig:d0_hww}b shows the
azimuthal opening angle distribution between electron and muon after all
cuts have been applied. This variable
has a strong discriminating power against, e.g. non-resonant $WW$
production, since leptons in signal events tend to be collinear due to
the spin correlations between the Higgs boson decay products.

Final data samples of $ee$, $e\mu$ and $\mu\mu$ final states contain
2, 2 and 5 events, respectively, while $2.7\pm0.4$, $3.1\pm0.3$ and
$5.3\pm0.6$ events are expected in MC. The background is dominated
by the $WW$ production. In the absence of a signal, 
the limits are set at $95\%$~C.L.
on the Higgs boson production cross section times branching ratio
into $W$ bosons as illustrated in Fig.~\ref{fig:d0_hww}c.

\section{Conclusions}

New measurements done by the CDF and D\O\ Collaborations for the SM
Higgs boson searches at the Tevatron Run~II, as well as the recent
improvements in accelerator performance are very encouraging.
Limits set on the $WH$ and
$H \ra WW^{(*)}$ production processes are unmatched or superior to
Run~I results.
To conclude -- the hunt for Higgs boson(s)~\footnote{Searches for non-SM
Higgs bosons at the Tevatron have been discussed in
a separate talk~\cite{nonsm_h}} in Run~II
has begun, and this searches will form a central part of the
physics program at the Tevatron.

\section*{Acknowledgments}
I am thankful to my colleagues from the CDF and D\O\ Collaborations
for providing material for this talk, and the Tevatron team for recent
significant improvements in accelerator performance. Special thanks
to the organizers of the Moriond 2004 Conference for
the stimulating atmosphere.

\end{document}